\documentclass[12pt]{iopart}
\usepackage{amsfonts,amssymb,amsmath}
\usepackage{subfigure,graphicx} 
\usepackage{flafter}
\usepackage{multirow}
\usepackage{txfonts}
\usepackage[unicode,breaklinks]{hyperref}
\usepackage{siunitx}
\newcommand{\papertitle}{Bubble nucleation in a cold spin 1 gas}
\hypersetup{
    unicode=true,
    plainpages=false,
    pdftitle={\papertitle},
    pdfauthor={Thomas P. Billam, Kate Brown, Ian G. Moss},
    pdfsubject={\papertitle},
    colorlinks=true,
    linkcolor=blue,
    citecolor=blue,
    filecolor=black,
    urlcolor=blue
}

\usepackage{ulem}

\newcommand{\be}{\begin{equation}}
\newcommand{\ee}{\end{equation}}
\newcommand{\bea}{\begin{eqnarray}}
\newcommand{\eea}{\end{eqnarray}}
\newcommand{\beal}{\begin{aligned}}
  \newcommand{\eeal}{\end{aligned}}

\begin{document} 

\title{\papertitle}

\author{Thomas P.\ Billam}
\ead{thomas.billam@ncl.ac.uk}
\address{Joint Quantum Centre (JQC) Durham--Newcastle, School of Mathematics, Statistics and Physics, 
Newcastle University, Newcastle upon Tyne, NE1 7RU, UK}

\author{Kate Brown}
\ead{k.brown@ncl.ac.uk}
\address{School of Mathematics, Statistics and Physics, 
Newcastle University, Newcastle upon Tyne, NE1 7RU, UK}

\author{Ian G. Moss}
\ead{ian.moss@ncl.ac.uk}
\address{School of Mathematics, Statistics and Physics, 
Newcastle University, Newcastle upon Tyne, NE1 7RU, UK}

\date{\today}

\begin{abstract}
Cold atomic gases offer the prospect of simulating the physics of the very early universe
in the laboratory. In the condensate phase, the gas is described by a field theory with 
key features of high energy particle theory. This paper describes a three level system which undergoes a first order phase transition through the nucleation of bubbles. The theoretical 
investigation shows bubbles nucleating in two dimensions at non-zero temperature. There is good agreement  between the bubble nucleation rates calculated from a Stochastic Projected Gross--Pitaevskii equation and from a non-perturbative instanton method. When an optical box trap is
included in the simulations, the bubbles nucleate preferentially near the walls of the trap.
\end{abstract}

\maketitle

\section{Introduction}

There has been speculation that the very early universe would have supercooled at various epochs into metastable phases, or even into ‘false vacuum’ states, before undergoing first order phase transitions. The ensuing violent fluctuations in
density would have echoes in the present day universe in the form of signals in
the cosmic microwave background \cite{PhysRevD.84.043507} and in a background of
gravitational waves \cite{Caprini:2009fx, Hindmarsh:2013xza}. This would likely have occurred at energies well above any that are accessible to experiment, and the phenomenon of false vacuum decay remains one of the most important yet untested phenomena in theoretical high energy particle physics.

The theoretical description of bubble nucleation at a first order transition devised in the 1970's involves an {\it instanton},
or \textit{bounce}, solution to the field equations in imaginary time
\cite{Coleman:1977py, Callan:1977pt, Coleman:1980aw}. 
However, the instanton approach gives limited information about how the bubbles emerge in real-time,
and how bubble nucleation events are correlated. 
A recent suggestion has been to fill the gaps in our understanding by exploring the details of false vacuum decay in ultracold atom systems, where the impressive
degree of experimental control available raises the
possibility of engineering (preferably quasi-relativistic) supercooled states and false vacua. The first scheme of this type due to Fialko et al.  \cite{FialkoFate2015,FialkoUniverse2017} concerns a two-component Bose gas in one dimension, formed from two spin states of a spinor condensate, coupled by a time-modulated microwave field. After time-averaging, one obtains an effective description containing a metastable false vacuum state in addition to the true vacuum ground state. An alternative proposal, from the present authors, uses a three-component condensate with Raman and RF mixing \cite{Billam:2021nbc}. In this paper we will present more details of this new scheme, and provide the first analysis of bubble nucleation for this system in two dimensions. 

Refs.~\cite{FialkoFate2015,FialkoUniverse2017,Braden:2017add,Billam:2018pvp,HertzbergQuantitative2020}
studied the decay of the false vacuum in the Fialko et al. scheme using field-theoretical instanton techniques and real-time simulations based on the truncated Wigner (TW) methodology~\cite{Steel1998,blakie_dynamics_2008}. The two descriptions appeared to align quite well. The scheme of Fialko et al. has also been extended to a finite-temperature 1D Bose gas~\cite{Billam:2020xna,Billam:2021psh,Ng:2020pxk}, with the aim of studying thermodynamical first order phase transitions in a cold atom system. Working in the time-averaged
effective description, both instanton techniques and the stochastic projected
Gross--Pitaevskii equation (SPGPE) \cite{GardinerStochastic2002, GardinerStochastic2003,bradley_bose-einstein_2008, BradleyStochastic2014} were used to investigate the decay of a supercooled gas which has been prepared in the
metastable state at low (but nonzero) temperatures. These methods showed excellent agreement in their predictions for the rate of the resulting first-order phase transition.

However, Refs.~\cite{Braden:2017add,Braden:2019vsw} showed that the false vacuum state in the Fialko et al. scheme can suffer from a parametric instability caused by the time-modulation of the system. This
instability presents a challenge to experimental implementation of the scheme
\cite{Braden:2017add,Braden:2019vsw,Billam:2020xna}. Furthermore, the scheme
requires inter-component interactions to be small compared to intra-component
interactions; this necessitates working very close to a Feshbach
resonance \cite{FialkoFate2015,FialkoUniverse2017}, which limits flexibility in
the experimental setup. The alternative scheme based on a spin 1 condensate is free from parametric instability and is more flexible in terms of experimental setup. Numerical simulations using the TW approximation in one dimension have demonstrated that the system undergoes vacuum decay in a way that is analogous to a Klein-Gordon system.

The one-dimensional systems give very limited information about realistic bubble nucleation events, yet more realistic three dimensional systems are difficult to probe experimentally. Two dimensions offer an ideal compromise for the discovery of important new phenomena, and in the present paper we will present a theoretical analysis of the nucleation of bubbles in a two dimensional (2D) version of the spin 1 system at finite temperature.

\section{System}

We will describe the system in two dimensions, assuming the atoms to be
tightly harmonically confined in the transverse direction such that a quasi-2D
description is suitable. We consider a condensate of alkali atoms in their $F=1$ hyperfine ground state manifold. The degeneracy between internal spin states $|m
\rangle$, where $m\in\{-1,0,1\}$, is lifted by a static magnetic field $B_z$
along the $z$ axis. In addition to intrinsic collisional coupling between the
spin states, described by a quartic Hamiltonian $H^{COL}$, we propose the states be
extrinsically coupled by both radio frequency fields (RF coupling) and by
optical fields in a two-photon Raman scheme (Raman coupling).

In this section we introduce the theoretical model, and the following sections describe the ground states and an example potential for atoms confined in an optical trap.

\begin{center}
\begin{figure}[htb]
  \centering
  \includegraphics[width=0.3\columnwidth]{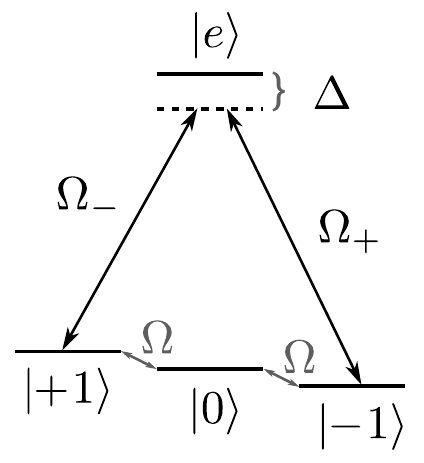}
\caption{Level coupling diagram for the simplest $\Lambda$ system based on
Raman and microwave induced transitions. The $F=1$ spin states labelled
$|m\rangle$, are coupled by a resonant RF beam of frequency $\omega_z$, with
Rabi frequency $\Omega$, and by a two-photon Raman coupling induced by
off-resonant optical beams with Rabi frequencies $\Omega_{\pm}$, zero
two-photon detuning, and detuning $\Delta$ from the excited state $| e
\rangle$.
\label{pot}}
\end{figure} 
\end{center} 

The terms in our mean field Hamiltonian are
\begin{equation}
H=\int d^2x\left\{
\overline\psi\left[\dfrac{-\hbar^2\nabla^2}{ 2m}-\mu\right]\psi+\overline\psi H^{ZE}_B\psi
+\overline\psi H^{MIX}_B\psi
\right\}+H^{COL},
\end{equation}
where the field $\psi$ has three components $\psi_m$.  The constant magnetic field
produces a first order Zeeman effect with frequency $\omega_z=g_F \mu_\mathrm{B} B_z / \hbar$ and a
second order Zeeman effect with frequency $\omega_q$,
\begin{equation}
H^{ZE}_B= \hbar\omega_z J_z+\hbar\omega_q\,J_z^2,
\end{equation}
where $J_x$, $J_y$ and $J_z$ are the dimensionless angular momentum generators.  The RF field
is tuned to the Zeeman frequency $\omega_z$ and is polarised in the $x$ direction.  
This directly couples
states with azimuthal quantum numbers $m\leftrightarrow m\pm1$.  Coupling of
the $m\leftrightarrow m\pm2$ states can be achieved by two optical fields
arranged on the $D_1$ line, creating a two-photon Raman coupling between the
states in a three-level $\Lambda$ scheme, as shown in Fig.~\ref{pot}. In
presenting our system we neglect complications arising from other states in the
upper hyperfine manifold, and consider only a single excited state $|e \rangle$
with azimuthal quantum number zero. To avoid population of $|e\rangle$, the
detuning $\Delta$ should be large compared to relevant atomic linewidths, and
to keep the momentum transferred to the atoms negligible the optical fields
driving $\sigma_\pm$ transitions should be co-propagating in the $z$-direction \cite{wright_raman}.
We assume zero two-photon detuning. After time-averaging of the RF and optical
frequencies in the rotating wave approximation as described in \ref{AppendixA}, 
we obtain the mixing part of
the Hamiltonian
\begin{equation}
H^{MIX}_B=\frac12\hbar\Omega J_x+\hbar\alpha \left(J_x^2+J_y^2\right),
\end{equation}
where the frequency $\Omega=g_F \mu_\mathrm{B} B_x / \hbar$ depends on the RF field amplitude $B_x$,
and $\alpha=-\Omega_+\Omega_-/4\Delta_e$ is determined by the optical-field Rabi
frequencies $\Omega_\pm$ and the detuning $\Delta_e$.

We assume that the atomic collisions in the Hamiltonian $H^{COL}$ are described by rotation invariant
dipole-dipole interactions $(\overline\psi\psi)^2$ and 
$(\overline\psi {\bf J}\,\psi)^2=(\overline\psi J_x\,\psi)^2+(\overline\psi J_y\,\psi)^2+(\overline\psi J_z\,\psi)^2$,  
which we would expect to describe a whole range of systems with
low to moderate external magnetic fields \cite{Kawaguchi2012,Stamper-Kurn2013}.
The interaction terms can be gathered together into an interaction potential
function $V_{\rm int}$, so that the total Hamiltonian becomes
\begin{equation}
H=\int d^2x\left\{
\overline\psi\left[\dfrac{-\hbar^2\nabla^2}{ 2m}\right]\psi+V_{\rm int}(\bar\psi,\psi)
\right\},
\end{equation}
where
\begin{align}
V_{\rm int}&=-\mu\overline\psi\psi+\hbar\omega_q(\overline\psi J_z^2\psi)
+\frac12g(\overline\psi\psi)^2+\frac12g'(\overline\psi {\bf J}\,\psi)^2\notag\\
&+\frac12\hbar\Omega\overline\psi J_x\psi+
\hbar\alpha\overline\psi \left(J_x^2+J_y^2\right)\psi.
\end{align}
The scattering parameters in the 2D system are
\begin{equation}
g=\left(\frac{8\pi\hbar^3\omega_\perp}{m}\right)^{1/2}\frac{a_0+2a_2}{3},
\qquad g'=\left(\frac{8\pi\hbar^3\omega_\perp}{m}\right)^{1/2}\frac{a_2-a_0}{3},
\end{equation}
where $a_F$ is the $s$-wave scattering length for total-spin-$F$
channels \cite{Kawaguchi2012,Stamper-Kurn2013}, and $\omega_\perp$ is the trap
frequency of the transverse confinement. Note that the linear Zeeman
term is cancelled out by the RF field in the rotating wave approximation. The appropriate
treatment of the spin-1 system for our purposes is one with a fixed chemical
potential but no additional Lagrange multiplier for the magnetisation, since the latter
is not conserved due to mixing between the spin states.

\section{Theoretical Analysis}

We shall show that the spin 1 system described above can have a meta-stable state which decays via bubble nucleation. Furthermore, the system is pseudo-relativistic, meaning that the system behaves dynamically like a relativistic system. It will prove convenient to re-scale the system to natural units. The healing length 
$\xi=\hbar/(mg\rho)^{1/2}$ and natural frequency $\omega_0=g\rho/\hbar$ are defined in terms of the density $\rho$ of one of the five phases listed below. We use the healing length as the length unit, $1/\omega_0$ as the time unit and $g\rho$ 
as the energy unit. Dimensionless parameters $\epsilon$ and $\lambda$ describe the strength of the mixing terms,
$\epsilon^2=\hbar\Omega/g\rho$ and $\lambda^2=\Omega_+\Omega_-/\Omega\Delta_e$.
We will continue to work in these units throughout the text, although we quote physical units in figure captions and when discussing experimental realizations.

\subsection{Ground states}

We begin with the phase structure in the absence of mixing terms.
Following Kawaguchi and Ueda \cite{Kawaguchi2012}, the fields can be parameterised by
\begin{align}
\psi_{\pm1}&=\sqrt{\rho}\,\zeta_{\pm1}\,e^{i(\theta\pm\varphi)},\label{param}\\
\psi_0&=\sqrt{\rho}\,\zeta_0,
\label{param2}
\end{align}
subject to $\zeta_0^2+\zeta_{+1}^2+\zeta_{-1}^2=1$.
The configuration space is made up of the quadrant $\zeta_0>0$, $\zeta_->0$
of the sphere and angular phases $0<\theta<\pi$, $0<\varphi<\pi$. 

\begin{table}[htb]
    \centering
\begin{tabular}{ |c|c|c|c|c| }
\hline
  \hspace{1em}phase\hspace{1em} & $\zeta_{+1}$ &  $\zeta_{-1}$ &  $\zeta_{0}$ & $\quad m_z \quad$\\ 
\hline
 F & 1 & 0 & 0 & 1\\
 F & 0 & 1 & 0 & 1\\
 AF & $1/\sqrt{2}$ & $1/\sqrt{2}$ & 0 & 0\\
 P & 0 & 0 & 1 & 0\\
 BA & $\displaystyle\ \frac{1}{2}\left(1 + \frac{g\omega_q}{2g'}\right)^{1/2}$\ & $\displaystyle\frac{1}{2}\left(1 + \frac{g\omega_q}{2g'}\right)^{1/2}$\ & $\displaystyle\ \frac{1}{\sqrt{2}}\left(1-\frac{g\omega_q}{2g'}\right)^{1/2}\  $ & 0 \\[12pt]
\hline
\end{tabular}
    \caption{Ground states of the spin 1 system with rotation 
    symmetric couplings and no mixing terms}
    \label{tab:phases}
\end{table}

Ferromagnetic phases (F) are characterised  by having magnetisation $m_z=\zeta_+^2-\zeta_-^2=\pm 1$.
The other phases are the antiferromagnetic (AF) phase with $\zeta_0=0$, 
the polar (P) phase with $\zeta_0=1$ and the broken axisymmetric phase (BA).
The values of the moduli $\zeta_i$ for zero magnetisation $m_z$ are shown
in Table \ref{tab:phases}. We single out the BA phase, which has the lowest energy when $g'<0$, $g>0$ and 
$0<g\omega_q<-2g'$. In the absence of mixing terms,
$\zeta_{+1}=\zeta_{-1}=\zeta_{BA}$ at the minimum (see Table \ref{tab:phases}).
Furthermore, we work in the regime $|g'/g| \ll 1$, where the chemical potential 
$\mu \approx g \rho$. 

\begin{center}
\begin{figure}[htb]
  \centering
\begin{center}
 \includegraphics[width=0.5\columnwidth]{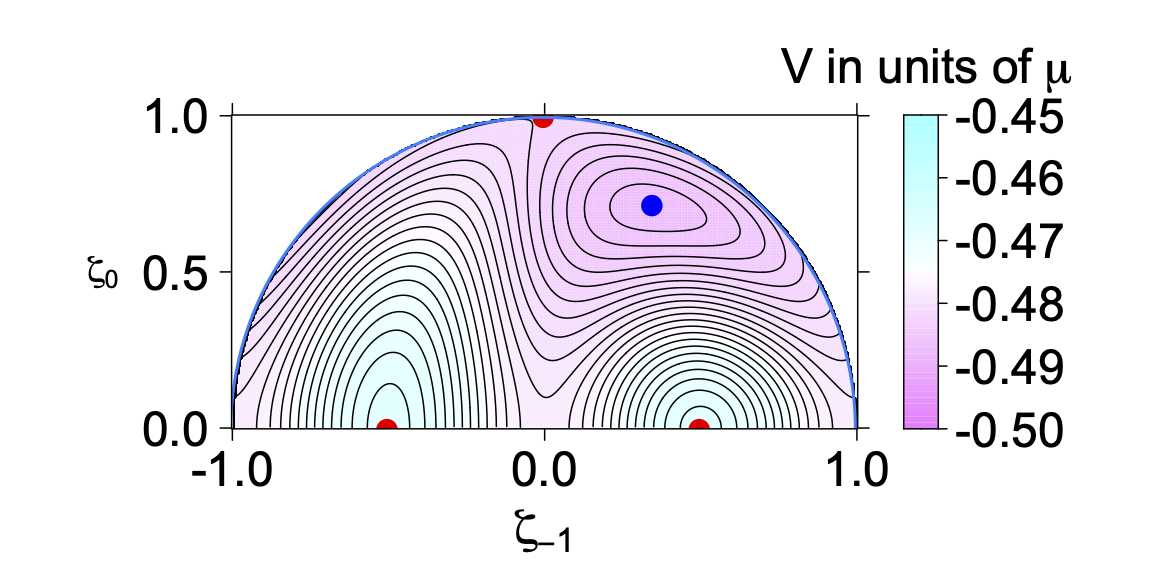}
 \includegraphics[width=0.4\columnwidth]{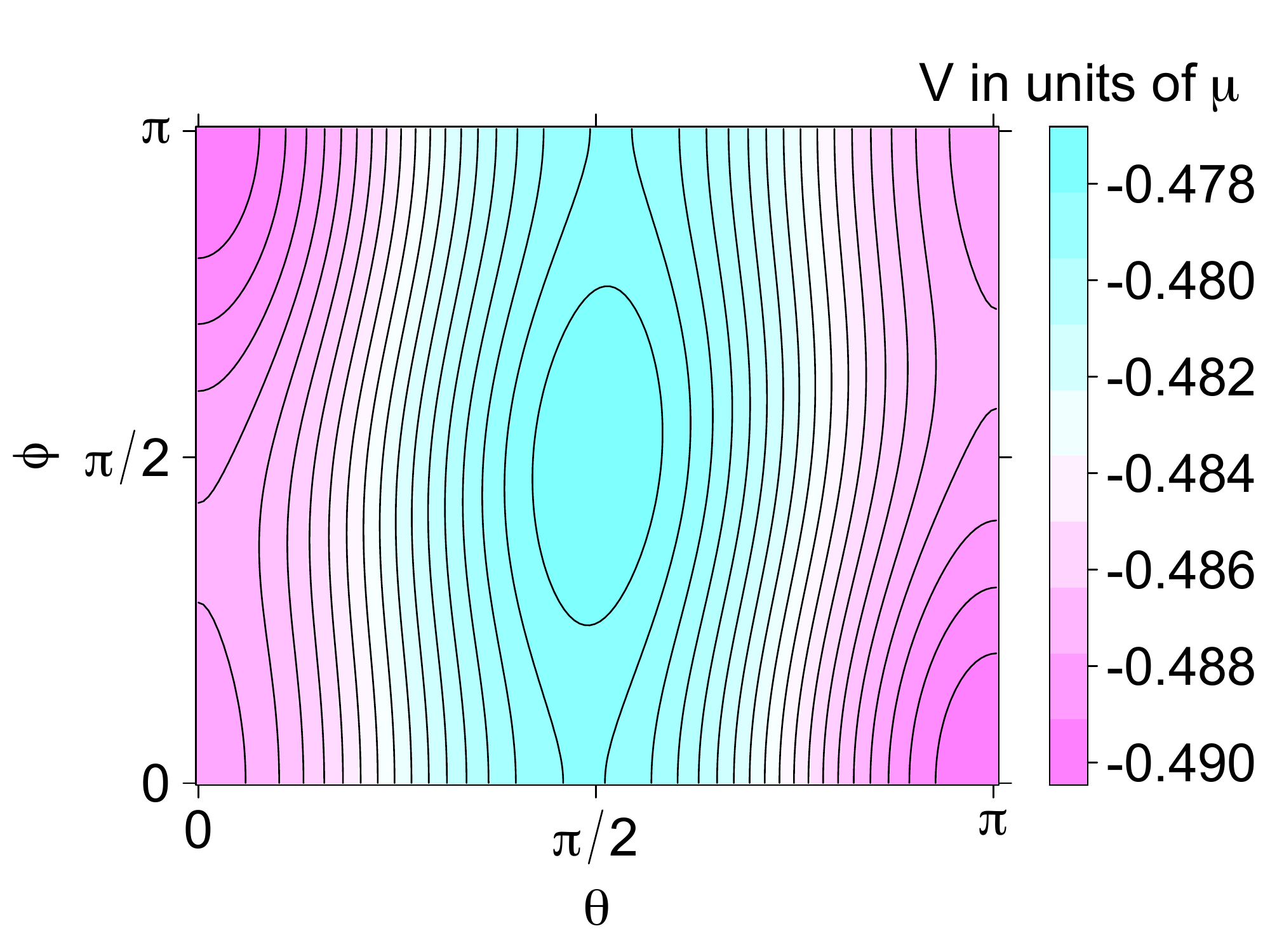}
\end{center}
\caption{The false vacuum, with RF and Raman mixing terms included,
is shown in different projections of the potential $V$ (in units of $\mu$).
The left-hand figure shows the potential as function of the relative amplitudes of the spin components at $(\theta,\varphi)=(0,0)$ and fixed $\mu$
using a Mollweide projection.
The BA vacuum state is indicated by a blue dot.
Right, the potential as a function of the phase angles at $\zeta_\pm=\zeta_{BA}$.
The false vacuum is at $(\theta,\varphi)=(0,0)$ and the true vacuum at $(\pi,0)$.
In this example, $g'=-0.0256g$, $\epsilon=0.05$, $\lambda=1.7$ and $\omega_q=0.017\mu/\hbar$
(see Table \ref{tab:constants}).
}
\label{vacV}
\end{figure}
\end{center}

In the regime of weak mixing, $\epsilon\ll 1$, the
states have approximately the same moduli as above. Crucially, however, the
weak mixing terms raise the degeneracy between different values of the phase so
that there are stationary points when $(\theta,\varphi)$ equals $(0,0)$,
$(\pi,0)$, $(0,\pi)$ and $(\pi,\pi)$.  The second derivatives of the potential
imply that the stationary points become local minima when
$\epsilon^2\lesssim -2g'/g$ and $\lambda \gtrsim 1$, as shown in the example
plotted in Fig. \ref{vacV}.

\subsection{Klein-Gordon limit}

The relevant sector of Bogliubov-de Gennes modes has dispersion relation
\begin{equation}
\omega(k)=\frac12\left(k^2+2\omega_q\right)^{1/2}\left(k^2+m_\varphi^2\right)^{1/2},
\end{equation}
where the effective mass $m_\varphi=O(\epsilon)$.
This reduces to a Klein-Gordon dispersion relation in the range $k\ll (2\omega_q)^{1/2}$. When combined with limits on the quadratic Zeeman shift
for the BA vacuum, fluctuations will appear relativistic when 
$k\ll 2|g'/g|^{1/2}$.
It follows that it is more difficult to replicate relativistic 
behaviour in systems with very small values of $|g'/g|$.

The Klein-Gordon mode can be isolated by fixing $\rho$ and taking 
$\zeta_\pm=\zeta e^{\pm\varphi/2}$.
The effective Lagrangian density ${\cal L}_{\rm eff}$ at  $O(\epsilon^2)$ then describes a Klein-Gordon field $\varphi$ with effective Lagrangian,
\begin{equation}
{\cal L}_{\rm eff}=2\zeta_{BA}^2\,\rho \left\{
\frac1{2c^2}(\partial_t\varphi)^2-\frac12(\nabla\varphi)^2
-V_{\rm eff}(\varphi)\right\}.
\end{equation}
The propagation speed of the Klein Gordon field is $c$, where $c^2=\omega_q/2$
in healing length units. The potential $V_{\rm eff}(\varphi)$ is
\begin{equation}
V_{\rm eff}=\epsilon^2\lambda_c^2\cos\varphi+\frac12\lambda^2\epsilon^2\sin^2\varphi,
\label{veff}
\end{equation}
where
\begin{equation}
\lambda_c=\left(\dfrac{1-g\omega_q/2g'}{
1+g\omega_q/2g'}\right)^{1/2}.
\end{equation}
Notice that the effective theory is a fully non-linear Klein-Gordon theory with non-polynomial interactions. The potential has a true vacuum at $\varphi=\pi$ and a false vacuum at $\varphi=0$ provided that $\lambda>\lambda_c$. 
The effective mass in the false vacuum state is $m_\varphi=\epsilon(\lambda^2-\lambda_c^2)^{1/2}$.

\subsection{Bubble nucleation}
\label{inst_predict}
The existence of a false vacuum state implies the possibility of supercooling. Small fluctuations about the false vacuum have insufficient energy to overcome the potential barrier around the state. Large fluctuations can eventually overcome the barrier, through the process of bubble nucleation. For a semi-classical model of bubble nucleation, we solve the field equations in imaginary time $\tau$ to give an instanton solution $\psi_b$. The instanton solution interpolates between the low energy state in the centre and the metastable state $\psi_{FV}$ at large distances. A slice through the instanton solution at $\tau=0$ represents a bubble. In the high temperature limit, the instanton is independent of $\tau$ and has rotational symmetry in space. 

The full expression for the 
nucleation rate of bubbles per unit area is
\cite{Coleman:1977py,Callan:1977pt},
\begin{equation}
\Gamma \approx A B[\psi_b]e^{-B[\psi_b]}.\label{gamma}
\end{equation}
where $B[\psi_b]$ denotes the difference in action between the instanton and the
metastable state, divided by $\hbar$. The pre-factor $A$ depends on the change in the spectra of the perturbative modes induced by the instanton. This should only depend
mildly on parameters, so we will treat this term as an undetermined constant.

The exponent is explicitly
\begin{equation}
B[\psi_b]=\rho\int d^2\mathbf{r} d\tau \left\{\psi_b^\dagger\dfrac{\partial \psi_b}{\partial\tau}
+\frac12\psi_b^\dagger\nabla^2\psi_b+V_{\rm int}(\psi_b)-V_{\rm int}(\psi_{FV})\right\}.
\end{equation}
In the Klein-Gordon approximation, the decay exponent simplifies to
\begin{equation}
B[\varphi_b]=2\zeta_{BA}^2\rho \int d^2\mathbf{r} d\tau\left\{
\frac{1}{2c^2}(\partial_\tau\varphi_b)^2+\frac12(\nabla\varphi_b)^2
+V_{\rm eff}(\varphi_b)-V_{\rm eff}(\varphi_{FV})
\right\},
\end{equation}
where $c=(\omega_q/2)^{1/2}$ and $\varphi_{FV}=0$. The potential was given in Eq. (\ref{veff}), with
\begin{equation}
\zeta_{BA}=\frac12\left(1+\frac{g\omega_q}{2g'}\right)^{1/2},\qquad
\lambda_c=\left(\dfrac{1-g\omega_q/2g'}{ 1+g\omega_q/2g'}\right)^{1/2}.
\end{equation}
The parameter dependence can be reduced if we make a change in coordinates,
\begin{equation}
\mathbf{r}'=\epsilon\lambda_c \mathbf{r},\qquad \tau'=c\epsilon\lambda_c \tau,\qquad \lambda'=\lambda/\lambda_c,
\qquad \chi=\dfrac{2\zeta_{BA}^2\rho}{ c\epsilon\lambda_c}.
\end{equation}
The exponent is recast into the form
\begin{equation}
B[\varphi_b]=\chi\int d^2\mathbf{r}'d\tau'\left\{\frac12(\partial_{\tau'}\varphi_b)^2
+\frac12(\nabla'\varphi_b)^2+V'(\varphi_b)-V'(0)\right\},
\end{equation}
where
\begin{equation}
V'(\varphi)=\cos\varphi+\frac12\lambda^{\prime 2}\sin^2\varphi.
\end{equation}
The numerical value for the exponent with this potential was calculated in Ref \cite{Abed:2020lcf}, with the result that $B=24\chi(\lambda'-1)/T'$. The temperature $T'$ scales like $1/\tau'$, hence
\begin{equation}
B=48\zeta_{BA}^2\frac{\rho}{T}\left(\dfrac{\lambda}{\lambda_c}-1\right).
\end{equation}
Converting the rate to the original coordinates gives
\begin{equation}
\Gamma=A'c(\epsilon\lambda_c)^3Be^{-B},
\label{eq:Gamma_inst}
\end{equation}
where the constant $A'$ only depends on $\lambda$.
In the numerical runs considered below, $\zeta_{BA}^2=1/6$, $\lambda=1.7$ and $\lambda_c=\sqrt{2}$
which give $B\approx 1.6\rho/T$.

\section{Numerical Investigation}
We perform numerical simulations using a \textit{simple growth} stochastic projected Gross-Pitaevskii equation (SPGPE) \cite{blakie_dynamics_2008, GardinerStochastic2002,
GardinerStochastic2003, bradley_bose-einstein_2008, BradleyStochastic2014}. In terms of our dimensionless variables, this is given by:
\begin{equation}
    i\frac{\partial {\psi_m}}{\partial {t}} = \mathcal{P}\Bigg\{(1-i\gamma)\Bigg[ -\frac{1}{2}{\nabla}^{2}{\psi}_m + \frac{\partial {V}}{\partial \psi_{m}^{\dagger}}\Bigg] + \eta_{m}\Bigg\}, \quad \quad m = -1, 0, 1,
\end{equation}
where we choose the Gaussian noise source $\eta$ to be uncorrelated between components, with correlations
\begin{equation}
\Big\langle \eta_{m}\big(\mathbf{r},t\big)\eta_{m'}^{\dagger}\big(\mathbf{r}',t'\big)\Big\rangle = 2\gamma T/\rho \hspace{0.1cm}\delta\big(\mathbf{r} - \mathbf{r}'\big)\delta\big(t - t'\big)\delta_{mm'}.
\end{equation} 
The projector $\mathcal{P}$ disregards modes with momentum $k^2 > k_{\textrm{cut}}^2$, where $k_{\textrm{cut}} = (2 T)^{1/2}$. This ensures that only modes that are sufficiently well described by the classical field approximation are included. This is similar to the approach used to investigate thermal bubble nucleation in a different atomic physics setup in Refs.~\cite{Billam:2020xna, Billam:2021psh}. Throughout this work, we fix the dimensionless dissipation rate at $\gamma = 0.02$.\\

\subsection{Periodic system}
Our baseline simulations consider a 2D system of size $L = L_x = L_y = 300$ with periodic boundaries and grid size of $\Delta l = \Delta x = \Delta y = 0.39$. The geometry of the potential term $V$ is set by fixing $\lambda = 1.7$ and $\epsilon = 0.1$.
The system was evolved using the fourth-order Runge-Kutta algorithm with time step $\Delta t = 0.04$ (agreement with tests at $\Delta t = 0.004$ proved the former to be sufficiently small). Our simulations were executed using the software package XMDS2~\cite{DennisXMDS2013}. Averaged quantities were calculated over a minimum of 100 stochastic realisations. We set the quadratic Zeeman shift to $\omega_q = -2g'/3g$ and consider \textsuperscript{7}Li, for which $g'/g = -0.456$.\\

In analogy with potential experiments, we initialize in a purely stable state. We allow the system to thermalise in the true vacuum until some time $t_{\text{switch}}$, about which the system is coerced into a metastable state, by means of a control parameter: 
\begin{equation}
    \alpha(t) = \dfrac{\pi}{2}\Bigg[1-\tanh\bigg( \dfrac{t - t_{\text{switch}}}{\tau_\mathrm{switch}} \bigg)\Bigg].
\end{equation}
Here, $\tau_\mathrm{switch}$ denotes the duration of the vacuum switch, which is implemented by modulating the RF-potential:
\begin{equation}
    V_{\text{RF}} = \dfrac{1}{2}\epsilon^2\psi^{\dagger}J_{x}\psi \rightarrow \frac12\cos(\alpha)\epsilon^2\psi^{\dagger}J_{x}\psi.
\end{equation}
In our simulations, we set $t_{\text{switch}} = 200$ and $\tau_\mathrm{switch} = 2.5$. \\

\begin{center}
\begin{figure}[htb]
  \centering
  \includegraphics[width=0.95\columnwidth]{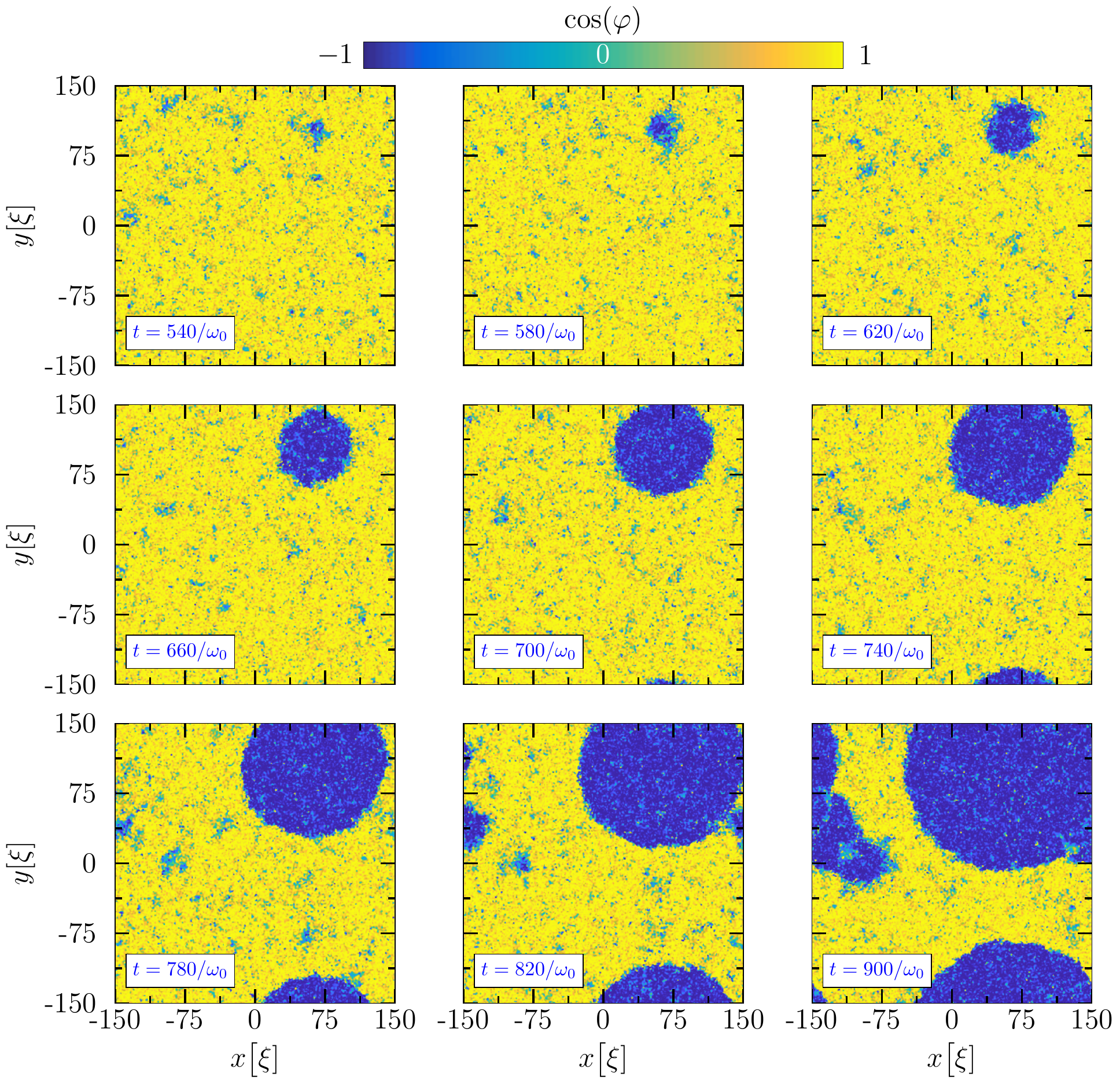}
\caption{An example realisation of bubble nucleation in \textsuperscript{7}Li with dimensionless coupling parameters $\lambda = 1.7$ and $\epsilon = 0.1$, density $\rho = 21 / \xi^2$ and temperature $T=2.4 \mu / k_\mathrm{B}$.}
\label{fig:bubbles}
\end{figure} 
\end{center}

\begin{center}
\begin{figure}[htb]
  \centering
  \includegraphics[width=0.4\columnwidth]{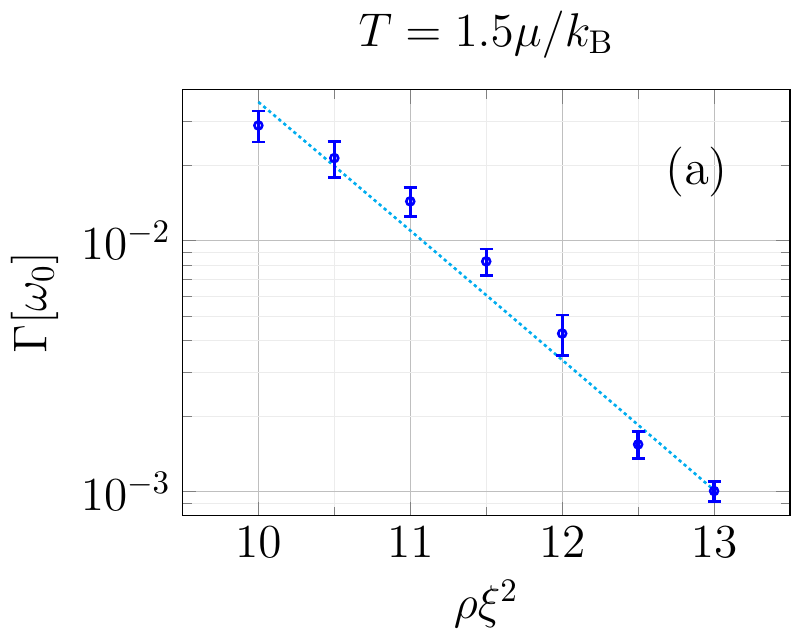}\hspace{1cm}
  \includegraphics[width=0.4\columnwidth]{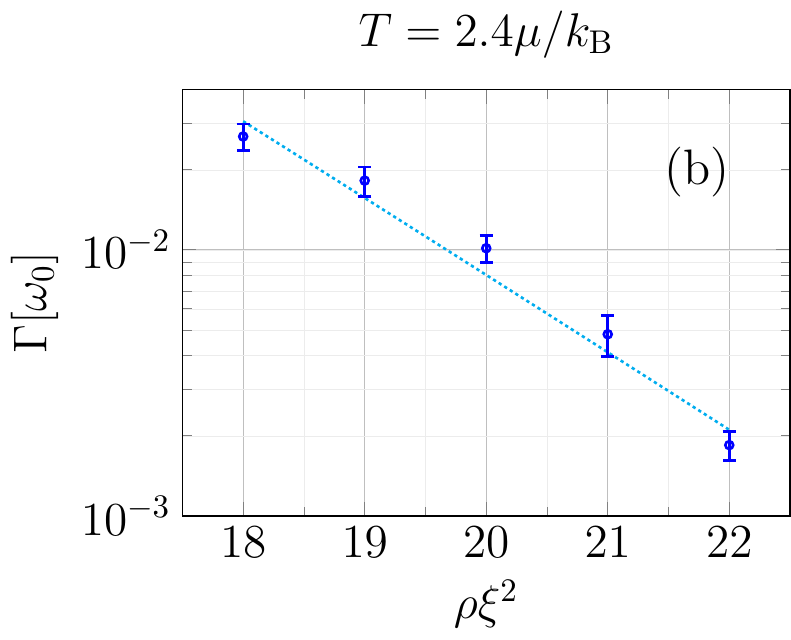}
  \includegraphics[width=0.4\columnwidth]{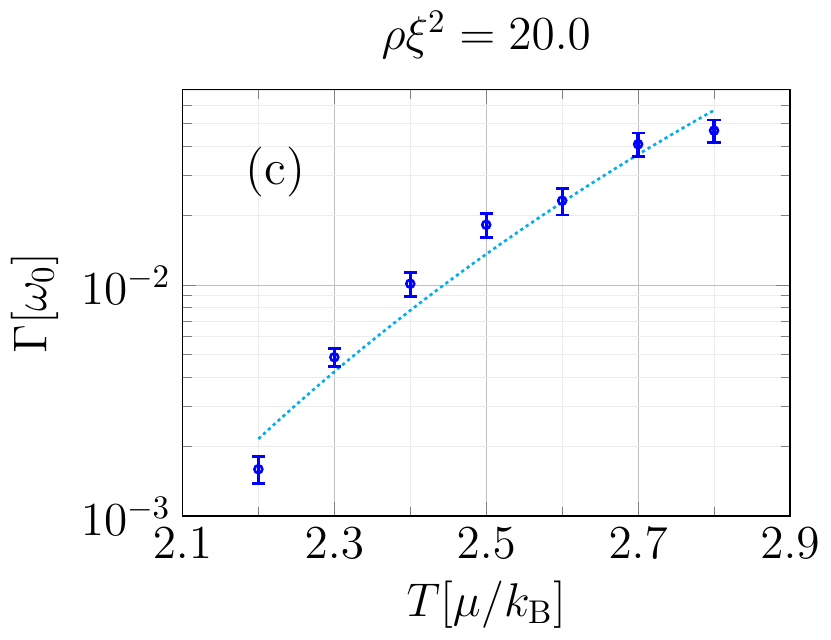}
\caption{The vacuum decay rate $\Gamma$ for \textsuperscript{7}Li, as a function of the density $\rho$ (made dimensionless as $\rho\xi^2$) (a - b), and of the temperature $T$ (c). In these plots $\lambda = 1.7$ and $\epsilon = 0.1$. Each panel includes a comparison between the SPGPE results (data points) and their instanton fit (dotted line). }
\label{fig:Gamma}
\end{figure} 
\end{center}

The evolution of $\cos{(\varphi)}$ is shown for a single stochastic realisation in Figure \ref{fig:bubbles}. This behaviour is typical; a bubble of true vacuum nucleates and expands roughly spherically. Further bubbles appear and collide with one another. Late snapshots hint that the likelihood of bubble nucleation may increase in the vicinity of a sufficiently large bubble. However, an investigation into this is beyond the scope of this work. We are primarily interested in the rate of false vacuum decay, $\Gamma$. To obtain this, we first examine the probability, $P$, of remaining in the false vacuum state. This is given by the proportion of stochastic trajectories which satisfy $\langle \cos(\phi)\rangle > 0.5$ at any time. We calculate $\Gamma$ by fitting $Ce^{-\Gamma t}$ to the exponentially-decaying region, $t_{\text{start}}\leq t \leq t_{\text{end}}$, where $t_{\text{start}}$ is the first time which satisfies $P(t)\leq 0.6$ and $t_{\text{end}} = t_0 + 0.1(t_0 - t_{\text{start}})$. Here, $t_0$ is the first time which satisfies $P(t) = 0$. This regime ensures that the fit termination depends consistently on the duration of decay. The dependence of $\Gamma$ on both temperature and density is explored in Figure \ref{fig:Gamma}. In line with the instanton prediction of Section~\ref{inst_predict}, we find that given a fixed temperature, the rate of vacuum decay decreases as density increases, whereas for fixed density, $\Gamma$ increases with temperature. Throughout this work, error bars are calculated using the bootstrap procedure detailed in \cite{Billam:2018pvp}. Here, we find the uncertainty in decay rate to be largest for the highest values of $\Gamma$, which can be attributed to a shift from first to second order behaviour. A more precise comparison with equation \eqref{eq:Gamma_inst} has been made by fitting $\Gamma_{\text{inst}} = a(\rho/T)\exp\big\{-b(\rho/T)\big\}$, where $a$ and $b$ are free to vary, to each simulated data curve. These fits are weighted by the error bars of the simulated data points. In general, we find good agreement between approaches. The fits for $b$ from Figure \ref{fig:Gamma} panels (a), (b), and (c) are $b= 1.9 \pm 0.3$, $b=1.7\pm0.4$, and $b=1.8\pm0.3$ respectively, where the uncertainty is quoted from the 95\% confidence interval. Although a systemic deviation in curve shape is arguably present, these values are remarkably consistent with the instanton prediction $b=1.6$.\\

\subsection{Trapped system}
\begin{center}
\begin{figure}[htb]
  \centering
  \includegraphics[width=0.7\columnwidth]{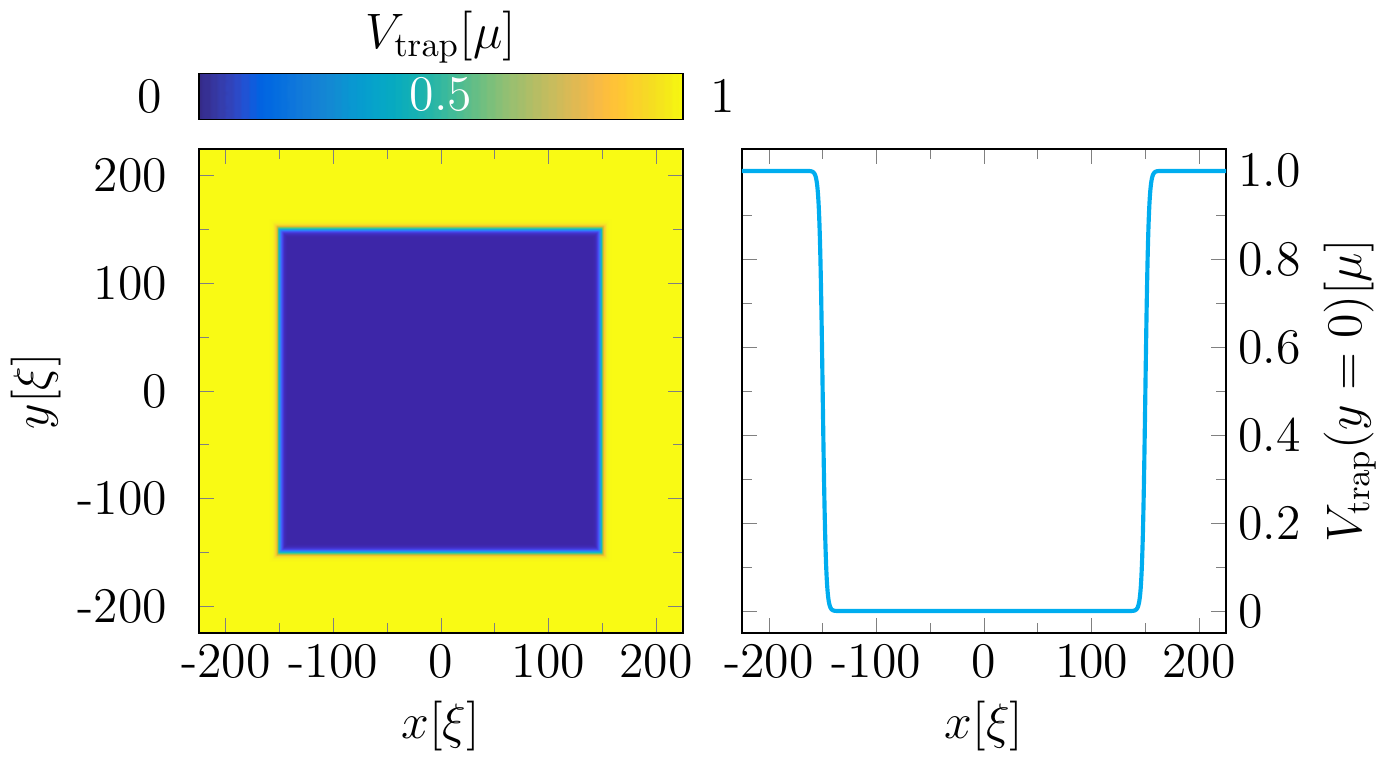}
\caption{The square trapping potential $V_{\text{trap}} = \max\Big\{V(x),V(y)\Big\}$, where $V(x)$ is defined in Equation \eqref{eq:trap1}. Here, $l_0 = 150\xi$ and $\sigma = 3\xi$. Left: The full spatial profile of $V_{\text{trap}}$. Right: a slice through $V_{\text{trap}}$ at $y = 0$.}
\label{fig:Vtrap}
\end{figure} 
\end{center}

In order to test the experimental viability of our investigations, we examined the effects of adding a trapping potential to the system. We proceeded with a periodic setup in the numerics, but increased the box size to $L_{\text{trap}} = (3/2)L$, whilst conserving $\Delta l$, and introduced a square trapping potential of the form $V_{\text{trap}} = \max\Big\{V(x),V(y)\Big\}$, where 
\begin{equation}
    V(x) = \frac{1}{2} \Bigg[2 + \tanh\bigg( \dfrac{x - l_0}{\sigma}\bigg) - \tanh\bigg( \dfrac{x + l_0}{\sigma}\bigg) \Bigg], \label{eq:trap1}
\end{equation}
Here, $l_0$ is the trap width and $\sigma$ is the trap wall thickness. Throughout this work, we fix $l_0 = (1/2)L$ and $\sigma = 3$; the former limiting the inhabitable region of the system to a box of same size as the un-trapped system. The potential $V_{\text{trap}}$ is shown for these parameters in Figure \ref{fig:Vtrap}. We also lowered the wall thickness to $\sigma = 1$, but this had negligible effect.

The inclusion of a trapping potential introduces a further complication; we no longer have analytic formulae for the vacuum states and must find these numerically. We first seek the Thomas-Fermi (TF) \cite{TF_ref} ground state corresponding to $\theta = \varphi =0$. This is found by making the transformation $\mu \rightarrow \mu - V_{\text{trap}}$ and solving the standard Gross-Pitaevskii equation (GPE) under the assumption that the kinetic and $\mathcal{O}(\epsilon^2)$ terms can be neglected. In the parameterisation \eqref{param}-\eqref{param2}, the standard Thomas-Fermi approximation becomes,
\begin{align}
    \rho_{\text{TF}} = \begin{cases}
                        \dfrac{\mu - V_{\text{trap}} - \omega_q/2}{g(1+g'/g)} \qquad &\textrm{if } \mu - V_{\text{trap}} - \omega_q/2 > 0,\\
                        \hspace{1.3cm} 0  & \textrm{otherwise,}\\
                       \end{cases}
    \label{eq_rho_TF}
\end{align}
We then proceed by propagating the Thomas-Fermi solution in real time, using a damped GPE:
\begin{equation}
    i\dfrac{\partial \psi_{m}}{\partial t} = (1 - i)\Bigg[ -\frac{1}{2}{\nabla}^{2}{\psi}_m + \frac{\partial {V}}{\partial \psi_{m}^{\dagger}}\Bigg],
    \label{eq:GPEdamped}
\end{equation}
where the chemical potential and trapping potential are included in $V$. The simulation is run for sufficient time to allow the wavefunction to converge to a stable vacuum state, which is then input as the initial conditions of the usual SPGPE procedure.\\

\begin{center}
\begin{figure}[htb]
  \centering
  \includegraphics[width=0.95\columnwidth]{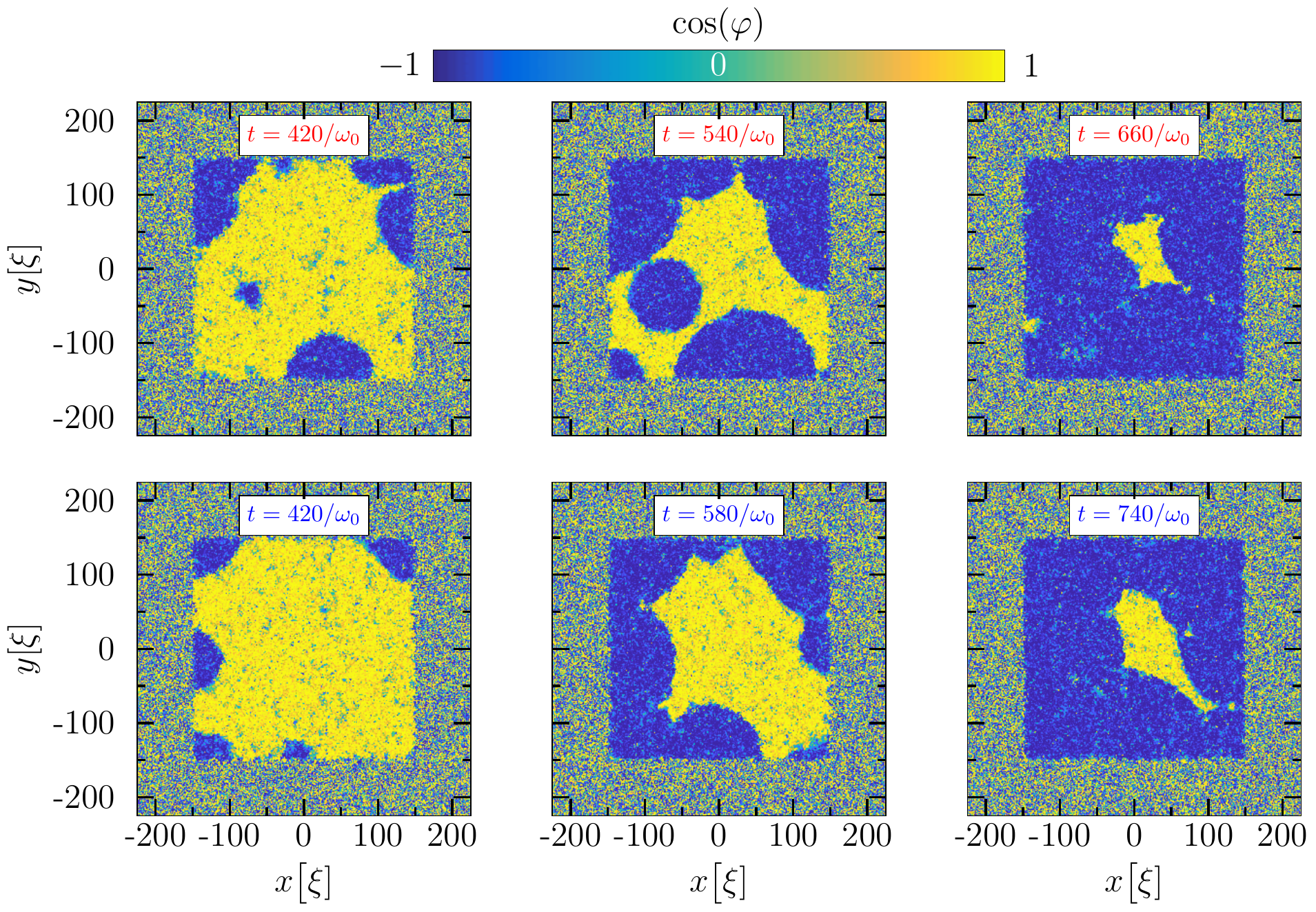}
\caption{The varying behaviour of example $\cos({\varphi})$ realisations in the presence of a square external potential, $V_{\text{trap}}$. Parameters are the same as in Figure \ref{fig:bubbles}, but with density $\rho = 20/\xi^2$. In all observed realisations, bubbles first nucleate around the trap boundaries. Top row (rare): A bubble nucleates away from the wall, which is eventually engulfed by the boundary bubbles. Bottom row (common): Boundary bubbles expand and fill the trap before a central bubble can form.}
\label{fig:barrier_bubbles}
\end{figure} 
\end{center}

\begin{center}
\begin{figure}[htb]
  \centering
  \includegraphics[width=0.4\columnwidth]{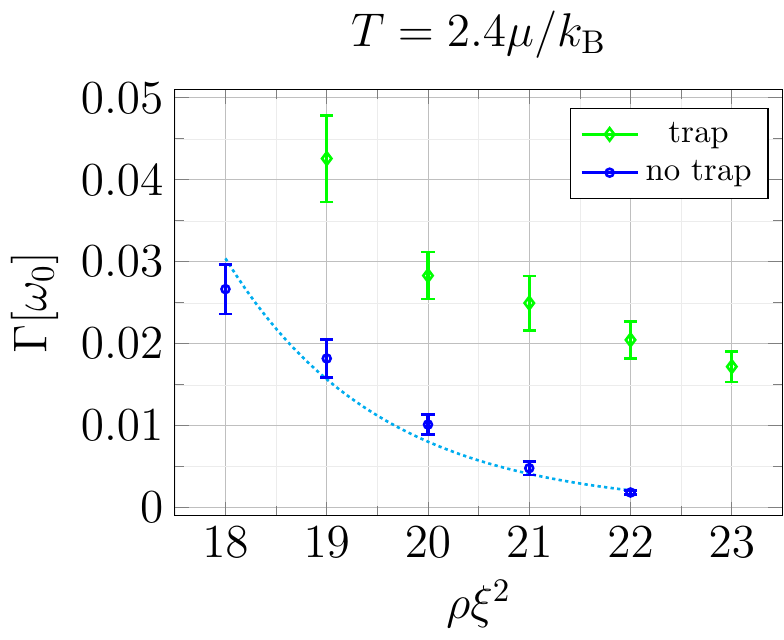}
\caption{A comparison of the vacuum decay rate $\Gamma$ in the presence (green diamond) and absence (blue circle) of $V_{\text{trap}}$. Here, $\Gamma$ is plotted as a function of the density $\rho$ (made dimensionless as $\rho \xi^2$), for $\lambda = 1.7$ and $\epsilon = 0.1$. The instanton fit (dotted line) is included alongside the simulated data for the trap-less case.}
\label{fig:barrier_Gamma}
\end{figure} 
\end{center}

The behaviour of $\cos(\varphi)$ in the presence of $V_{\text{trap}}$ is explored in Figure \ref{fig:barrier_bubbles}. The addition of boundaries accelerates the bubble nucleation process; the trap walls themselves act as nucleation sites. In general, bubbles form along these first, before expanding to enclose and ultimately fill the inhabitable region. Bubbles rarely have time to form away from the walls, and any such bubbles are eventually consumed by their older neighbours. An example of this is included in Figure \ref{fig:barrier_bubbles}. In order to increase the yield of central bubbles, and prolong their existence, we suggest increasing the system size substantially. Due to computational cost, we refrained from doing this. The effect of boundaries is made more explicit in Figure \ref{fig:barrier_Gamma}, where vacuum decay is plotted as a function of density for both the trapped and un-trapped systems. The inclusion of $V_{\textrm{{trap}}}$ induces a global increase in $\Gamma$ over the whole range of density values investigated.

\section{Experimental Realisation}

The relevant physical properties for alkali species with the required property 
$g'/g< 0$ are tabulated in Table~\ref{tab:constants}. The ground state hyperfine 
energy splitting $\Delta E_\mathrm{hfs}$ determines the magnetic field needed 
to achieve a given
quadratic Zeeman shift~\cite{Stamper-Kurn2013}. While $g'/g$ is fixed by the
atomic species, there is considerable flexibility in choosing tunable
experimental parameters that correspond to the dimensionless parameters used in
our simulations. As an example, a system with parameters similar to those used in Figs. \ref{fig:Gamma}, but with a larger density $\rho = 50/\xi^2$ would correspond 
to around $50000$ $^7$Li
atoms in a $\SI{100}{\micro m}$ wide square optical trap with transverse
frequency $\omega_r = 2\pi \times \SI{50}{\kilo Hz}$ and a bias field of
$B_z=0.27\,$Gauss.  The timescale $\omega_0^{-1}$ corresponds to
$\SI{1.1}{\milli s}$. Such a system would have a smaller extent ($32\xi$) than our simulations when measured in healing lengths, and the temperature unit $\mu / k_\mathrm{B} \approx \SI{7}{\nano K}$ would be small in comparison to $\hbar \omega_\perp / (2k_\mathrm{B})$.
An alternative scheme with potassium
would correspond to around $1.9\times10^6$ $^{41}$K atoms in a $\SI{170}{\micro
m}$ wide square trap with transverse
frequency $\omega_r = 2\pi \times \SI{1.9}{\kilo Hz}$ and a bias field
of $B_z=0.08\,$Gauss. Such a system would have timescale $\omega_0^{-1} = \SI{0.2}{\milli s}$, density $\rho = 21 / \xi^2$, and an extent of $300\xi$ similar to our simulations. The energy scales would satisfy $\mu \lesssim \hbar \omega_\perp/2 \lesssim k_\mathrm{B}T$ when the temperature is a few times the temperature unit $\mu / k_\mathrm{B} \approx \SI{37}{\nano K}$. In any experiment, we assume there is very wide experimental flexibility in
terms of the coupling field Rabi frequencies and detuning ($\Omega$,
$\Omega_\pm$, $\Delta$); in practice these would need to be tuned to give the
desired $\epsilon$ and $\lambda$ by taking into account the additional,
smaller, light shifts arising from the other states in the upper hyperfine
manifold that we neglect here.
Finally, we note that while we have described a system with RF mixing between the $1,0$ and the $0,1$ levels, 
the proposal should work equally well if these levels are coupled by Raman transitions instead.

\begin{table}
\centering
\begin{tabular}{lcccl}
Species & $a_0\,(a_\mathrm{Bohr})$ & $a_2\,(a_\mathrm{Bohr})$ & $g'/g$ & $\Delta E_\mathrm{hfs}\,$(MHz) \\
\hline  \\ [-1.5ex]
$^{7}$Li & $23.9$ & $6.9$ & $-0.456$ & $803.5 \times h$ \cite{Li_hyperfine_ref} \\
$^{41}$K & $68.5$ & $63.5$ & $-0.0256$ & $254.0 \times h$ \cite{K_hyperfine_ref_1, K_hyperfine_ref_2}  \\
$^{87}$Rb & $101.8$ & $100.4$ & $-0.0046$ & $6834.7 \times h$ \cite{Rb_hyperfine_ref}
\end{tabular}
\caption{\label{tab:constants}Physical properties used to compute simulation parameters. Scattering lengths are from the table in Ref.~\cite{Stamper-Kurn2013}.}
\end{table}

\section{Conclusion}

We have proposed that an optically Raman coupled spin-1 Bose gas
can be used to simulate first order phase transitions and bubble nucleation 
in a Klein-Gordon system. Unlike previous proposals, our system is
free from resonant instability and avoids the need to reduce inter-component
scattering lengths using Feshbach resonances. 

The system can be used to investigate the nucleation of bubbles in real time
and the effects of quantum coherence between different bubbles \cite{Pirvu:2021roq}. 
Such details are difficult to model using computer simulations, so in this paper
we have analysed the thermal case where a stochastic approach is known to be reliable.
We found that the bubble nucleation rate shows good agreement
with instanton methods, and we expect the simulations give a genuine real-time picture of the nucleation process. In the vacuum case, an alternative approach,
such as the Truncated Wigner (TW) method, 
needs to be employed. The use of the TW method for false vacuum decay
has been investigated in Refs. \cite{Braden:2018tky,Braden:2022odm}.
The validity of the TW method is something we plan to 
explore in future work.

Data supporting this publication are openly available under
a Creative Commons CC-BY-4.0 License in \cite{data_package}.

\appendix

\section{Combined Raman and RF Mixing}\label{AppendixA}

In this appendix we describe how the combination of optical and radio frequency
beams leads to mixing between the spin states. The physics is based on a three-level
$\Lambda$ scheme for the Raman coupling, with a new extension to include the extra 
RF mixing.

A constant magnetic field $B_z$ is applied along the $z$ axis,
for linear Zeeman splitting $\hbar\omega_z$, with frequency $\omega_z=g_F\mu_B B_z/\hbar$.
The RF mixing is provided by a beam along the $z$ axis with frequency $\omega_z$, and 
Hamiltonian $H^{RF}=-\mu_B{\bf J}\cdot{\bf B}$, where ${\bf B}$ is polarised along the $x$ axis,
\begin{equation}
{\bf B}=\frac12{\bf e}_x{\cal B}e^{-i\omega_z t}+\hbox{c.c.}
\end{equation}
The optical beams are also applied along the $z$ axis and couple to the
dipole moment ${\bf d}$ of the wave function, with Hamiltonian $H^{RAM}=-{\bf d}\cdot{\bf E}$, where
the electric field is
\begin{equation}
{\bf E}=\frac12\boldsymbol{\epsilon}_+{\cal E}_+e^{-i\omega_+t}
+\frac12\boldsymbol{\epsilon}_-{\cal E}_-e^{-i\omega_-t}+\hbox{c.c.}
\end{equation}
The dipole strengths are $d_{nm}^\pm=\langle n|{\bf d}\cdot\boldsymbol{\epsilon}^\pm |m\rangle$,
with the polarisation chosen so that $d_{3e}^-=d_{1e}^+=0$.

In the state basis $|1\rangle,|2\rangle,|3\rangle,|e\rangle$, the total mixing Hamiltonian is
\begin{equation}
H_M=
\begin{pmatrix}
\hbar\omega_z&V_{12}&0&V_{1e}\\
V_{21}&0&V_{23}&0\\
0&V_{32}&-\hbar\omega_z&V_{3e}\\
V_{e1}&0&V_{e3}&E_e\\
\end{pmatrix}
\end{equation}
with RF mixing matrix elements
\begin{equation}
V_{12}=V_{23}=-\frac{1}{2\sqrt{2}}\mu_B {\cal B}e^{-i\omega_z t}+\hbox{c.c}
\end{equation}
and optical mixing matrix elements
\begin{align}
V_{1e}&=-\frac12d_{1e}^-{\cal E}^-e^{-i\omega_-t}+\hbox{c.c}\\
V_{3e}&=-\frac12d_{3e}^+{\cal E}^+e^{-i\omega_+t}+\hbox{c.c}
\end{align}

In order to implement the rotating wave approximation we make a 
change of basis, $\Psi_n=\psi_ne^{-i\omega_n t}$.
In the $\psi_n$ basis,
\begin{equation}
H_M=
\begin{pmatrix}
\hbar(\omega_z-\omega_1)&V_{12}e^{i(\omega_1-\omega_2)t}&0&V_{1e}e^{i(\omega_1-\omega_e)t}\\
V_{21}e^{i(\omega_2-\omega_1)t}&-\hbar\omega_2&V_{23}e^{i(\omega_2-\omega_3)t}&0\\
0&V_{32}e^{i(\omega_3-\omega_2)t}&-\hbar(\omega_z+\omega_3)&V_{3e}e^{i(\omega_3-\omega_e)t}\\
V_{e1}e^{i(\omega_e-\omega_1)t}&0&V_{e3}e^{i(\omega_e-\omega_3)t}&E_e-\hbar\omega_e\\
\end{pmatrix}
\end{equation}
A simple regime occurs when we choose $\omega_z=(\omega_+-\omega_-)/2$, and
\begin{align}
\omega_1&=\omega_z\\
\omega_2&=0\\
\omega_3&=-\omega_z\\
\omega_e&=(\omega_++\omega_-)/2
\end{align}
The Hamiltonian becomes
\begin{equation}
H_M=
\begin{pmatrix}
0&V_{12}e^{i\omega_0t}&0&V_{1e}e^{-i\omega_-t}\\
V_{21}e^{-i\omega_0t}&0&V_{23}e^{i\omega_0t}&0\\
0&V_{32}e^{-i\omega_0t}&0&V_{3e}e^{-i\omega_+t}\\
V_{e1}e^{i\omega_-t}&0&V_{e3}e^{i\omega_+t}&\hbar\Delta_e\\
\end{pmatrix}
\end{equation}
where $\Delta_e$ is the excited state detuning,
\begin{equation}
\hbar\Delta_e=E_e-\hbar(\omega_++\omega_-)/2
\end{equation}
On timescales longer than $1/\omega_z$, we average to get
\begin{equation}
H_{RWA}=\dfrac{\hbar}{2}
\begin{pmatrix}
0&\Omega/\sqrt{2}&0&\Omega_-^*\\
\Omega^*/\sqrt{2}&0&\Omega/\sqrt{2}&0\\
0&\Omega^*/\sqrt{2}&0&\Omega_+^*\\
\Omega_+&0&\Omega_-&2\Delta_e\\
\end{pmatrix}
\end{equation}
where $\hbar\Omega=-\mu_B{\cal B}$, and the Rabi frequencies are
\begin{align}
\hbar\Omega_-&=-d_{e1}^-{\cal E}^-\\
\hbar\Omega_+&=-d_{e3}^+{\cal E}^+
\end{align}
Adiabatic elimination of the excited state (setting $\dot \psi_e=0$) gives
\begin{equation}
H_\Lambda=\dfrac{\hbar}{2}
\begin{pmatrix}
\displaystyle -\dfrac{|\Omega_-|^2}{2\Delta_e}&\Omega/\sqrt{2}&
\displaystyle -\dfrac{\Omega_-\Omega_+^*}{2\Delta_e}\\[12pt]
\Omega^*/\sqrt{2}&0&\Omega/\sqrt{2}\\[6pt]
\displaystyle -\dfrac{\Omega_-^*\Omega_+}{2\Delta_e}&\Omega^*/\sqrt{2}&
\displaystyle -\dfrac{|\Omega_+|^2}{2\Delta_e}\\
\end{pmatrix}
\end{equation}
The diagonal terms can be absorbed into (or can replace) the quadratic Zeeman term.
In operator form
\begin{equation}
H_\Lambda=\frac12\hbar\Omega{\bf J}_x+\hbar\alpha({\bf J}_x^2+{\bf J}_y^2)
\end{equation}
where $\alpha=-\Omega_-\Omega_+/4\Delta_e$ when $\Omega$ and $\Omega_\pm$ are real.\\

Data supporting this publication is openly available under a Creative Commons CC-BY-4.0 License in Ref. \cite{data_package}

\section*{Acknowledgements}
This work was supported in part by the Science and Technology Facilities Council (STFC) [grant ST/T000708/1] and the UK Quantum Technologies for Fundamental Physics programme [grants ST/T00584X/1 and ST/W006162/1]. 
KB is supported by an STFC studentship. This research made use of the Rocket High Performance 
Computing service at Newcastle University.

\section*{References}
\bibliographystyle{iopart-num}
\bibliography{paper}

\providecommand{\newblock}{}
\begin{thebibliography}{10}
\expandafter\ifx\csname url\endcsname\relax
  \def\url#1{{\tt #1}}\fi
\expandafter\ifx\csname urlprefix\endcsname\relax\def\urlprefix{URL }\fi
\providecommand{\eprint}[2][]{\url{#2}}

\bibitem{PhysRevD.84.043507}
Feeney S~M, Johnson M~C, Mortlock D~J and Peiris H~V 2011 {\em Phys. Rev. D\/}
  {\bf 84}(4) 043507 (\textit{Preprint} \eprint{1012.3667})

\bibitem{Caprini:2009fx}
Caprini C, Durrer R, Konstandin T and Servant G 2009 {\em Phys. Rev. D\/} {\bf
  79} 083519 (\textit{Preprint} \eprint{0901.1661})

\bibitem{Hindmarsh:2013xza}
Hindmarsh M, Huber S~J, Rummukainen K and Weir D~J 2014 {\em Phys. Rev.
  Lett.\/} {\bf 112} 041301 (\textit{Preprint} \eprint{1304.2433})

\bibitem{Coleman:1977py}
Coleman S~R 1977 {\em Phys. Rev. D\/} {\bf 15} 2929--2936 [Erratum: Phys. Rev.
  D 16, 1248 (1977)]

\bibitem{Callan:1977pt}
Callan C~G and Coleman S~R 1977 {\em Phys. Rev. D\/} {\bf 16} 1762--1768

\bibitem{Coleman:1980aw}
Coleman S~R and De~Luccia F 1980 {\em Phys. Rev. D\/} {\bf 21} 3305

\bibitem{FialkoFate2015}
{Fialko} O, {Opanchuk} B, {Sidorov} A~I, {Drummond} P~D and {Brand} J 2015 {\em
  EPL (Europhysics Letters)\/} {\bf 110} 56001 (\textit{Preprint}
  \eprint{1408.1163})

\bibitem{FialkoUniverse2017}
{Fialko} O, {Opanchuk} B, {Sidorov} A~I, {Drummond} P~D and {Brand} J 2017 {\em
  Journal of Physics B Atomic Molecular Physics\/} {\bf 50} 024003
  (\textit{Preprint} \eprint{1607.01460})

\bibitem{Billam:2021nbc}
Billam T~P, Brown K and Moss I~G 2022 {\em Phys. Rev. A\/} {\bf 105} L041301
  (\textit{Preprint} \eprint{2108.05740})

\bibitem{Braden:2017add}
Braden J, Johnson M~C, Peiris H~V and Weinfurtner S 2018 {\em JHEP\/} {\bf 07}
  014 (\textit{Preprint} \eprint{1712.02356})

\bibitem{Billam:2018pvp}
Billam T~P, Gregory R, Michel F and Moss I~G 2019 {\em Phys. Rev. D\/} {\bf
  100} 065016 (\textit{Preprint} \eprint{1811.09169})

\bibitem{HertzbergQuantitative2020}
Hertzberg M~P, Rompineve F and Shah N 2020 {\em Phys. Rev. D\/} {\bf 102}
  076003 (\textit{Preprint} \eprint{2009.00017})

\bibitem{Steel1998}
Steel M~J, Olsen M~K, Plimak L~I, Drummond P~D, Tan S~M, Collett M~J, Walls D~F
  and Graham R 1998 {\em Phys. Rev. A\/} {\bf 58} 4824--4835 (\textit{Preprint}
  \eprint{cond-mat/9807349})

\bibitem{blakie_dynamics_2008}
Blakie P, Bradley A, Davis M, Ballagh R and Gardiner C 2008 {\em Advances in
  Physics\/} {\bf 57} 363 (\textit{Preprint} \eprint{0809.1487})

\bibitem{Billam:2020xna}
Billam T~P, Brown K and Moss I~G 2020 {\em Phys. Rev. A\/} {\bf 102} 043324
  (\textit{Preprint} \eprint{2006.09820})

\bibitem{Billam:2021psh}
Billam T~P, Brown K, Groszek A~J and Moss I~G 2021 {\em Phys. Rev. A\/} {\bf
  104} 053309 (\textit{Preprint} \eprint{2104.07428})

\bibitem{Ng:2020pxk}
Ng K~L, Opanchuk B, Thenabadu M, Reid M and Drummond P~D 2021 {\em PRX
  Quantum\/} {\bf 2} 010350 (\textit{Preprint} \eprint{2010.08665})

\bibitem{GardinerStochastic2002}
Gardiner C~W, Anglin J~R and Fudge T~I~A 2002 {\em Journal of Physics B:
  Atomic, Molecular and Optical Physics\/} {\bf 35} 1555--1582
  (\textit{Preprint} \eprint{0112129})

\bibitem{GardinerStochastic2003}
Gardiner C~W and Davis M~J 2003 {\em Journal of Physics B: Atomic, Molecular
  and Optical Physics\/} {\bf 36} 4731--4753 (\textit{Preprint}
  \eprint{0308044})

\bibitem{bradley_bose-einstein_2008}
Bradley A~S, Gardiner C~W and Davis M~J 2008 {\em Phys. Rev. A\/} {\bf 77}
  033616 (\textit{Preprint} \eprint{0712.3436})

\bibitem{BradleyStochastic2014}
Bradley A~S and Blakie P~B 2014 {\em Phys. Rev. A\/} {\bf 90}(2) 023631
  (\textit{Preprint} \eprint{1406.2029})

\bibitem{Braden:2019vsw}
Braden J, Johnson M~C, Peiris H~V, Pontzen A and Weinfurtner S 2019 {\em
  JHEP\/} {\bf 10} 174 (\textit{Preprint} \eprint{1904.07873})

\bibitem{wright_raman}
Wright K~C, Leslie L~S and Bigelow N~P 2008 {\em Phys. Rev. A\/} {\bf 78}
  053412

\bibitem{Kawaguchi2012}
Kawaguchi Y and Ueda M 2012 {\em Physics Reports\/} {\bf 520} 253
  (\textit{Preprint} \eprint{1001.2072})

\bibitem{Stamper-Kurn2013}
Stamper-Kurn D~M and Ueda M 2013 {\em Rev. Mod. Phys.\/} {\bf 85}(3) 1191
  (\textit{Preprint} \eprint{1205.1888})

\bibitem{Abed:2020lcf}
Abed M~G and Moss I~G 2020  (\textit{Preprint} \eprint{2006.06289})

\bibitem{DennisXMDS2013}
Dennis G~R, Hope J~J and Johnsson M~T 2013 {\em Computer Physics
  Communications\/} {\bf 184} 201 (\textit{Preprint} \eprint{1204.4255})

\bibitem{TF_ref}
Pitaevskii L and Stringari S 2016 {\em {Bose-Einstein Condensation and
  Superfluidity}\/} International Series of Monographs on Physics (Oxford: OUP)
  ISBN 9780198758884

\bibitem{Li_hyperfine_ref}
Foot C 2005 {\em {Atomic Physics}\/} Oxford Master Series in Physics (Oxford:
  OUP) ISBN 9780198506966

\bibitem{K_hyperfine_ref_1}
Falke S, Tiemann E, Lisdat C, Schnatz H and Grosche G 2006 {\em Phys. Rev. A\/}
  {\bf 74} 032503

\bibitem{K_hyperfine_ref_2}
Arimondo E, Inguscio M and Violino P 1977 {\em Rev. Mod. Phys.\/} {\bf 49} 31

\bibitem{Rb_hyperfine_ref}
Bize S, Sortais Y, Santos M~S, Mandache C, Clairon A and Salomon C 1999 {\em
  Europhysics Letters ({EPL})\/} {\bf 45} 558

\bibitem{Pirvu:2021roq}
Pirvu D, Braden J and Johnson M~C 2022 {\em Phys. Rev. D\/} {\bf 105} 043510
  (\textit{Preprint} \eprint{2109.04496})

\bibitem{Braden:2018tky}
Braden J, Johnson M~C, Peiris H~V, Pontzen A and Weinfurtner S 2019 {\em Phys.
  Rev. Lett.\/} {\bf 123} 031601 [Erratum: Phys.Rev.Lett. 129, 059901 (2022)]
  (\textit{Preprint} \eprint{1806.06069})

\bibitem{Braden:2022odm}
Braden J, Johnson M~C, Peiris H~V, Pontzen A and Weinfurtner S 2022
  (\textit{Preprint} \eprint{2204.11867})

\bibitem{data_package}
Billam T~P, Brown K and Moss I~G 2022 {Data supporting publication: Bubble
  nucleation in a cold spin 1 gas} (Dataset available at
  doi:10.25405/data.ncl.21681809)

\end{thebibliography}
\end{document}